\documentstyle[twoside,fleqn]{article}

\newcommand{\bls}[1]{\renewcommand{\baselinestretch}{#1}}

\def\noi{\noindent}

\makeatletter
\renewcommand{\section}{\@startsection{section}{1}{0pt}%
        {-3.5ex plus -1ex minus -.2ex}{2.3ex plus .2ex}%
        {\large\bf\protect\raggedright}}

\renewcommand{\subsection}{\@startsection{subsection}{2}{0pt}%
        {-3ex plus -1ex minus -.2ex}{1.4ex plus .2ex}%
        {\normalsize\bf\protect\raggedright}}

\renewcommand{\thesubsubsection}%
        {\arabic{section}.\arabic{subsection}.\arabic{subsubsection}.}
\renewcommand{\@oddhead}{\raisebox{0pt}[\headheight][0pt]{%
   \vbox{\hbox to\textwidth{\rightmark \hfil \rm \thepage \strut}\hrule}}}
\renewcommand{\@evenhead}{\raisebox{0pt}[\headheight][0pt]{%
   \vbox{\hbox to\textwidth{\thepage \hfil \leftmark \strut}\hrule}}}
\newcommand{\heads}[2]{\markboth{\protect\small\it #1}{\protect\small\it #2}}

\makeatother

\newcommand{\Title}[1]{\noindent {\Large #1} \\}
\newcommand{\Author}[2]{\noindent{\large\bf #1}\\[2ex]\noindent{\it #2}\\}
\newcommand{\Abstract}[1]{\vskip 2mm \begin{center}
     \parbox{16.4cm}{\small\noindent #1} \end{center}\bigskip}

\def\nqq{\hspace{-2em}}

\newcommand{\sequ}[1]{\setcounter{equation}{#1}}

\def\beq{\begin{equation}}
\def\eeq{\end{equation}}
\def\bear{\begin{eqnarray}}
\def\al{&\nhq}
\def\lal{&&\nqq {}}               % left alignment
\def\bearr{\begin{eqnarray} \lal}
\def\ear{\end{eqnarray}}
\def\earn{\nonumber \end{eqnarray}}

\def\tst{\textstyle}

\def\yy{\\[5pt]}

\def\e{{\,\rm e}}

\def\half{{\tst\frac{1}{2}}}

\def\ve{\varepsilon}

\def\vp{\varphi}
\def\al{\alpha}
\def\b{\beta}
\def\k{\kappa}
\def\L{\Lambda}
\def\sr{\sqrt}

\def\th{\theta}

\def\ve{\varepsilon}

\def\e{\epsilon}
\def\l{\lambda}

\def\pt{\partial}
\newcommand{\Tr}{{\bf Tr}}
\heads{S.V.Chervon}
{Gravitational Field of the Early Universe: I.Non-linear Scalar Field 
as the Source.}

\bls{1.01}

\begin{document}
\thispagestyle{empty}
\twocolumn[
\noi \unitlength=1mm
\begin{picture}(174,8)
   \put(31,8){\shortstack[c]
       {RUSSIAN GRAVITATIONAL SOCIETY                \\
       ULYANOVSK STATE UNIVERSITY        }       }
\end{picture}
\begin{flushright}
                                         RGS-USU-97/01 \\
                                         gr-qc/9706028    \\
%	{\it Grav. and Cosmol.} {\bf 2}, 3\foom 1, 221-226 (1996)
\end{flushright}
\bigskip

\Title{GRAVITATIONAL FIELD OF THE EARLY UNIVERSE:\yy
        I.NON-LINEAR SCALAR FIELD AS THE SOURCE}

\Author{S.V.Chervon}
%\email{chervon@themp.univ.simbirsk.su}
  {Ulyanovsk State University,
  42 Leo Tolstoy Str., Ulyanovsk 432700, Russia}
  
{\it Received 20 March 1997} \\

\Abstract{
In this review article we consider three most important sources of the
gravitational field of the Early Universe: self-interacting scalar field,
chiral field and gauge field.
The correspondence between all of them are pointed out.
More attention is payed to nonlinear scalar field
source of gravity. The progress in finding the exact solutions in
inflationary universe is reviewed.
The basic idea of `fine turning of the potential'
method is discussed and computational background is presented in
details. A set
of new exact solutions for standard inflationary model and conformally-flat
space-times are obtained. Special attention payed to relations between
`fine turning of the potential' and Barrow's approaches.
As the example of a synthesis of both methods new exact solution is obtained.
%In this contribution it will be considered further development of
%the method of `fine turning of the potential'.
}
\centerline {\small PACS: 98.80+12.25}

\bigskip

%%%%%%%%%%%%%%%%%%%%%%%%%%%%%%%%%%%%%%%%%%%%%%%%%%%%%%%%%%%%%%%%%%%%%%%%%%%%%%%
]                        %% end `onecolumn'

\section{Introduction}   %% Sec.1
One of the most promising recent achievement in the physics of
the Early Universe is Inflationary Cosmology.
The inflationary Universe became now an integral part of
the hot Big Bang scenario and solved its long standing problems
such as horizon, flatness, homogeneity \cite{guth81}-\cite{albste82}.
%and large scale structure formation
%\cite{linde90},\cite{olive90}.

To describe the gravitational field of the very Early Universe, in general
case, we can consider the model with the action
\beq\label{1.1}
S=\int \sr{-g}d^4x \lbrace {\cal L}_G+{\cal L}_M+{\cal L}_{Int} \rbrace.
\eeq
where ${\cal L}_G =\frac{R+2\L}{2\k}, {\cal L}_{Int} $ describes the
interaction of a gravitational field to matter, ${\cal L}_M$ being
the lagrangian density of matter. Henceforth we put
$ {\cal L}_{Int}=0 $.

It was realized \cite{guth81},\cite{linde82} that during inflationary stage
the source of the gravitational field was gauge and Higgs fields described
by the Grand Unified Theory (GUT).
This model can be described by the the lagrangian
%(ref{0.0}) where
\sequ{1}
%\beq\label{1.1}
%{\cal S}= \int \sr{-g} d^4x \lbrace {\cal L}_G + {\cal L}_M \rbrace
%\eeq
%where
\beq\label{1.2}
{\cal L}_M=-\frac{1}{\gamma}\lbrace\frac{1}{2}G_{\mu\nu}^a G^{a\mu\nu}+
 D_{\alpha}\varphi^a D^{\alpha}\varphi^a + V(\vert\varphi\vert)\rbrace
\eeq
In the relation (\ref{1.2})
 $G_{\mu\nu}^{a} = \partial_{\mu}W^{a}_{\nu} -
 \partial_{\nu}W^{a}_{\mu} + ef^{a}_{bc}W^{b}W^{c} , f^{a}_{bc}$
 being a structure's constants of a gauge group for the Yang-Mills field
$W_{\mu}^{a},~ D_{\mu}\vp^a =\partial_{\mu}\vp^a +ef^a_{bc}W_\mu^b\vp^c $
being the covariant derivative,
 $\varphi^{a}$ being the Higgs field.
The potential can be taken in the form $V(\vert\varphi\vert) =
 \frac{\lambda}{4}(\varphi^{a}\varphi^{a} - F^2)^2$ , where
 $ F = \langle\vert\varphi^a\vert\rangle\not=0$ being the vacuum
average of the field.

The action (\ref{1.1})-(\ref{1.2}) leads to the system of
Einstein-Yang-Mills-Higgs equations.
To solve this self-consistent system is the real problem
even under assumption about the symmetry. Therefore in inflationary
scenarios  it was used an effective model of a scalar field
$\phi $ with the
potential of self-interaction $V(\phi)$ for the sake of simplicity
\cite{linde90}. The action
for `standard inflationary model' reads
\beq\label{1.3}
{\cal S}=\int \sr{-g}d^4x \lbrace {\cal L}_G +
\half \phi_{,i} \phi_{,k} g^{ik} - V(\phi)\rbrace.
\eeq
Recently it was proposed so-called chiral inflationary model
based on a nonlinear sigma model (NSM) with the
self-interacting potential \cite{ch95gc}. The action for this model can
be presented as
\beq\label{1.4}
{\cal S}= \int\limits_{\cal M} \sr{-g} d^4x \lbrace {\cal L}_G+ \frac{1}{2}
 h_{AB}(\vp) \vp^A_{,i} \vp^B_{,k} g^{ik} - W(\vp) \rbrace
\eeq
Here $({\cal M}, g_{ik}(x))$ being  a space-time,
 $({\cal N}, h_{AB}(\varphi))$
 being a target space, $\varphi = (\varphi^{1},...,\varphi^{n}),
~\varphi_k^A=\partial_k \varphi^A=\varphi^A_{,k}.$

The model (\ref{1.4}) can be considered as generalisation of standard
inflationary model (\ref{1.3}) and in special case can be reduced to it.
Namely, let us introduce new scalar field $\Phi $ and a potential
$\tilde V(\Phi )$ which satisfy the relations
\beq\label{1.4-1}
\Phi_{,i}\Phi_{,k}=h_{AB}\vp^A_{,i}\vp^B_{,k}
\eeq
\beq\label{1.4-2}
\tilde V(\Phi)=W(\vp^C)
\eeq
To realize the reduction above we have to restrict the metric of the chiral
space $h_{AB}$. Differentiating by coordinates $x^i$ and $x^k$ the relation
(\ref{1.4-2}) and then using (\ref{1.4-1}) one can obtain the restriction
on $h_{AB}$:
\beq\label{1.4-3}
h_{AB}=\frac{\pt W}{\pt \vp^A}\frac{\pt W}{\pt \vp^B} \Bigl(
\frac{\pt\tilde V}{\pt\Phi}\Bigr)^{-2},~ \frac{\pt\tilde V}{\pt\Phi}\ne 0.
\eeq
Thus the chiral inflationary model (\ref{1.4}) can be reduced to the standard
inflationary one (\ref{1.3}) by virtue of the relations (\ref{1.4-1})-
 (\ref{1.4-2}).

We can obtain as well the relation between GUTs
(\ref{1.1})-(\ref{1.2}) and chiral inflationary model (\ref{1.4}).
The model (\ref{1.4}) can be
obtained from (\ref{1.1})-(\ref{1.2}), if we put $W_\mu=0 $ and
$h_{AB}=diag[1,1,\ldots,1] $. The last expression means that the target
space  are postulated to be an eucludian one. Thus the chiral inflationary
model does not only simple reduction from GUT but it is the generalisation
of a Higgs sector by virtue of the change the eucludian space by riemannian
target space. It worth note one more interesting feature of the
chiral inflationary model (\ref{1.4}). Namely the field equations
\beq\label{1.5}
{1\over\sqrt {|g|}}\partial_i(\sqrt {|g|} \varphi_{A}^{,i})-
 \Gamma_{C,AB} \varphi_{,i}^B \varphi_{,k}^C g^{ik}-
{\partial W \over \partial \varphi^A} = 0
\eeq
contain an additional term (the second one) which came from the intrinsic
geometry of the target space. In standard inflationary models (\ref{1.3})
based on self-interacting scalar field theory a similar term
have been inserted by fenomenological way from quantum properties
\cite{brandenberger89}.
Moreover it is well known the direct relation between the
NSM with special symmetry and gauge theories (see, for example,
\cite{svv96iv}).

The exponential expansion of the Universe
has been found by G.Ivanov \cite{ivanov80} for
a nonlinear scalar field with the potential
$V(\phi)=\frac{\mu}{2}\phi^2-\frac{\l}{4}\phi^4 $
in the framework of spatially-flat Friedmann-Robertson-Walker (FRW) spaces
and has been interpretated as the Universe started from a quasi-vacuum state
of matter. Ivanov's solution has been obtained before the work by A.Guth
\cite{guth81} where the inflationary solution has been obtained with
$V(\phi)=const$.

Exact solutions of the power law inflationary type have been
obtained for Liouville non-linearity $V(\phi)=me^{-\l\phi}$ \cite{bbl86}
(see also \cite{barsai93} and references quoted therein).
The Liouville-type and some other non-linearities have been investigated
and some exact and asymptotic solutions are presented in \cite{muslimov90}.
Classical de Sitter type solutions was obtained in \cite{ellmad91}.
New classes of exact solutions have been found in \cite{barrow94} by
taking the scalar field as the function of time $\phi = \phi (t)$ and then
determining the evolution of the expansion scale factor $K(t)$ and the
potential $V(\phi (t))$ from it. This approach was applied in works
\cite{barpar95},\cite{parbar95} as well.

Nevertheless approximations and numerical investigations are used in a
large number of inflationary scenarios because of the difficulties in
obtaining the exact solutions for inflationary models.
The slow-roll approximation is the most common in use \cite{olive90}.

In  \cite{chezhu95},\cite{chezhu96iv} the exact solutions have been obtained
by taking, first, the scalar factor as the function of time $K=K(t)$
and then determining the evolution of the potential $V=V(t)$ and the
evolution of a scalar field $\phi = \phi (t)$, the dependence between
$V$ and $\phi$ being, in general, parametric one. This approach
called as the method of `fine turning of the potential' \cite{chezhu96iv}.

Another point have been presented in \cite{salbon90}, where exact general
solutions were found for a single scalar field interacting through an
exponential potential in the framework of background field equations
for the Arnowitt-Deser-Misner (ADM) formalism. Approximate analytic
solutions for slowly evolving multiple scalar fields are obtained also.
The Hamilton-Jacobi theory for long-wavelength inhomogeneous universes
are investigated in \cite{salopek91} in the framework of ADM formalism.
Exact inhomogeneous solutions for Yang-Mills field minimally coupled to
gravity have been obtained in \cite{szc96}.

In this review article we give the basic idea of a new insight on the
potential of self-interaction.
Using further development of `fine turning of the potential' method
\cite{chezhu96iv} we obtain new exact solutions in the case of the
conformally-flat spaces.
The difficulties of obtaining the exact solutions for
self-gravitating massive scalar field are pointed out.
The set of generalised Barrow's solutions \cite{barrow94}
is obtained.
\section{The effective self-interaction potential}
The original inflationary scenario \cite{guth81} as well as its first
modifications \cite{linde82},\cite{albste82} are connected with the effective
theory of self-interacting scalar field $\phi$, minimally coupled to
gravity, with the action (\ref{1.3})

As a rule the standard inflationary model (\ref{1.3}) is analysed
in the framework of the Friedmann-Robertson-Walker metric
\beq\label{2.1}
     dS^2=dt^2-K^2(t)\biggl(\frac{dr^2}{1-\e {r^2}}
     +r^2(d\th^2+\sin^2\th d\vp^2)\biggr).
\eeq
Here $\e=-1,0,+1$ corresponds to open, spatially-flat and closed Universe
respectively.

A very important role in inflationary scenarios belongs to the effective
potential of self-interaction $V(\phi)$.
The form of the $V(\phi)$ reflects the physical phenomenon at the very early
Universe: cosmological phase transitions and the symmetry restoration at
high temperatures $T$ in GUTs. It is well known that the form of a
potential
is changed while phase transitions occur and a temperature increase
\cite{linde90},\cite{olive90}. The potential depends on the temperature
and this dependence is due to quantum one-loop corrections in finite temperature
field theory.

Thus the form of the effective scalar potential depends on
the type of a field theory, we put into a physical basis, and tends to change
when physical phenomenon occur in the development of a cosmological time $t$.

One more restriction on the form of the potential $V(\phi,T)$ came from the
fine turning procedure \cite{shavil84}. Let us consider as an example the
situation with the Coleman-Weinberg potential \cite{colwei73}
\beq\label{2.2}
V(\phi )=A \phi^4 \{\ln (\frac{\phi^2}{M^2}+c)\},
\eeq
predicted by the $SU(5)$ GUT, and the fine turning of the potential
(\ref{2.2}).
Here $M$ is an arbitrary renormalization mass and $á$ is a constant of order
unity.
It is happened that the potential (\ref{2.2}) is unsuitable for matching the
$SU(5)$-
invariant GUT (with the $A\propto e^4 > 10^{-2}$) with the observed rate
of the density perturbations $\frac{\delta\rho}{\rho}$ (which needs
$A < 10^{-12}$). To improve the situation Shafi and Vilenkin \cite{shavil84}
took into consideration the three-level potential
$$
V(\phi ,\Phi ,H_5)=\half a(\Tr \Phi^2)^2 +\half b\Tr \Phi^4+
\al(H_5^{+}H_5)\Tr \Phi^2+
$$
\beq
\frac{1}{4}\l (H_5^{+}H_5)^2+
\frac{1}{4}\l_1\vp^4 -\half \l_2 \vp^2 \Tr \Phi^2+\l_3\phi^2 H_5^{+}H_5,
\eeq
containing the adjoint and fundamental Higgs fields
$\Phi $ and $H_5$ in addition to the inflationary field $\phi$.
It should be mentioned here, that in \cite{shavil84} the kinetic terms
from $\Phi$ and $H_5$
did not include in the model. This shortage has been avoided
in the multiple  \cite{star85}
and double-field inflationary models \cite{tvvsj87}.
 Let us also mention the concrete high energy physics phenomenon
which use double-field model with the potential, containing
cross-interaction \cite{sakkhl93} $\half \nu \vp^2\xi^2$.
\bear\label{2.4}
  &&V(\vp ,\xi )=\half m_\vp^2\vp^2+\frac{1}{4}\l_\vp \vp^4-\\
\nonumber
  &&\half m_\xi^2 \xi^2+\frac{1}{4} \l_\xi \xi^4 +\half \nu \vp^2\xi^2.
\ear

Thus we can see a tendency of changing the shape of the potential
$V(\phi,T)$ if one needs to correct the inflationary scenario.

Summarising above we can come to the conclusion, that the form of the
effective potential $V(\phi )$ does not fix and is subjected to change
with the evolution of the Universe.
Put into basis possible variations of the $V(\phi(t))$ we can present
a new insight on the problem of obtaining the exact solutions for
inflationary models.
Namely, what kind of the $V(\phi(t))$ admits
the exact solutions with an exponential or power law expansion of the FRW
Universe?
The answer to this question
will give us an explicit form of the
potential which leads to the given rate of the expansion of the Universe.
The next step should be to find an appropriate theory of
matter which will be more close to the obtained form of the potential.
We will not discuss the last problem here and will pay attention
to the way of construction the $V(\phi)$ in the next section.
\section{The method}
The system of Einstein's and nonlinear scalar field equations, corresponding to
the model (\ref{2.1}) in the FRW spaces (\ref{2.2}), reads
\bear\label{3.1}
  &&\frac{K_{44}}{K}+\frac{2K_{4}^2}{K^2}+\frac{2\epsilon}{K^2}
  =-\Lambda+\kappa{V(\phi)},\\ \label{3.2}
  &&-\frac{3K_{44}}{K}=\Lambda+\kappa(\phi^2_4-{V(\phi)}),\\ \label{3.3}
  &&\phi_{44}+3\frac{K_{4}}{K}\phi_4+\frac{dV(\phi)}{d\phi}=0.
\ear
These equations present the standard inflationary model.

To obtain new class of exact
solutions we will use a freedom in the choice of a potential's form.

Considering the equations (\ref{3.1})-(\ref{3.3}) one can find that
the last equation (\ref{3.3}) is the differential consequences of
(\ref{3.1}) and (\ref{3.2}). Really, to prove this fact, one can
differentiate (\ref{3.1}) by $t$ to obtain the following equation
\beq\label{3.3-1}
\frac{K_{444}}{K}+3\frac{K_{4}K_{44}}{K^2}-\frac{4K_4^3}{K^3}
-\frac{4K_4\e}{K^3}-\k \frac{dV}{dt}=0.
\eeq
This equation can be rewritten as
\bear\label{3.3-2}
&&\frac{K_{444}}{K}-3\frac{K_{4}K_{44}}{K^2}+\frac{2K_4^3}{K^3}
+\frac{2K_4\e}{K^3}- \\
&& -6\frac{K_4}{K}\left( -\frac{K_{44}}{K}+
\frac{K^2_4}{K^2}+\frac{\e}{K^2}\right)-\k \frac{dV}{dt}=0.
\ear
Now one can use two consequences of the
Einstein's equations (\ref{3.1}) and (\ref{3.2}):
\begin{itemize}
\item
the sum of Einstein's equations (\ref{3.1}) and (\ref{3.2}), firstly
\beq\label{3.4}
\k \phi_4^2=\frac{2}{K^2}\lbrace -K K_{44}+K_4^2+\e\rbrace ,
\eeq
\item
and, secondly, the derivative by time $t$ from the relation (\ref{3.4})
\beq\label{3.5}
\k \phi_{44}\phi_4=\pt_4\lbrace -\frac{K_{44}}{K}+\frac{K_4^2}{K^2}+
\frac{\e}{K^2}\rbrace
\eeq
\end{itemize}
Inserting left hand side of (\ref{3.4}) and (\ref{3.5}) into (\ref{3.3-2})
and using $\frac{dV}{dt} = \frac{dV}{d\phi} \phi_4 $,
we can divide the equations (\ref{3.3-2}) by $\phi_4 \ne 0$.
As the result one obtains the equation for the scalar field (\ref{3.3}).

The following analysis will be used just for Einstein's equations
(\ref{3.1}) and (\ref{3.2}), which can be reduced to the form
where the functions
$V(t)\equiv{V(\phi(t))}$ and $\phi(t)$
are expressed through the function
$K(t)$ and their derivatives \cite{chezhu96iv}:
\bear\label{3.6}
  &&V(t)=\frac{1}{\kappa}\left(\Lambda+\frac{K_{44}}{K}
  +\frac{2K_{4}^2}{K^2}+\frac{2\epsilon}{K^2}\right),\\
\label{3.7}
&&\phi(t)=\pm\sqrt{\frac{2}{\kappa}}\int\left(\sqrt{-
\frac{d^2\ln{K}}{dt^2}+
\frac{\e}{K^2}}\right) dt.
%+ \phi_0,
% \frac{d^2\ln{K}}{dt^2}%
% \frac{K_{44}}{K}+\frac{K_4^2}{K^2}+
% +\frac{\epsilon}{K^2}}\right)dt+\phi_0,
\ear
%where $\phi_0$ is a constant of integration.

By giving the rate of the expansion as the function for a scale factor
 on time $K=K(t)$ , we can find the functions
$\phi(t)$ ¨ $V(t)$ which are necessary for chosen type of the Universe's
evolution.
It is obvious, that the pair of the function (\ref{3.6}) and (\ref{3.7})
gives the parametric dependence
$V=V(\phi)$.
In some cases, after calculation of the right hand sides in
 (\ref{3.6}),
(\ref{3.7}), it is possible to find the explicit dependence
$V=V(\phi)$ by eliminating $t$.
\section{The inflationary solutions}

The exact solutions for exponential and power law type of inflation
in the framework
of `fine turning of the potential' (FTP) method have been obtained in
\cite{chezhu95,chezhu96iv,ch96el}.

\subsection{Power law inflation}
To obtain the power law inflation let us start from the suggestion
that
\beq\label{4.1} %3.8
K(t)=K_0t^m,
\eeq
where $m>1$.

The integral in the right hand side of
(\ref{3.7}) can be calculated explicitly.
Therefore we can find  $(m\not=1)$
\bear\label{4.2} %{EqVt}
&&V=\frac{m}{\kappa t^2}\left(\frac{\Lambda t^2}{m}-(3m-1-2\alpha{t}^{-2m+2})\right)\\
\nonumber %label{4.3} %{EqPhit}
&&\phi(t)=\pm\sqrt{\frac{1}{2\kappa}}\frac{m}{1-m}\left\{2\sqrt{1+\alpha{t}^{-2m+2}}\right.+\\
\label{4.3}
&&\left. +{\rm ln}\left(\frac{\sqrt{1+\alpha{t}^{-2m+2}}+1}{\sqrt{1+\alpha{t}^{-2m+2}}-1}\right)\right\}+\phi_0,
\ear
where $\alpha=\e {K_0^{-2}}/m$.

In the case of the spatially-flat
Universe ($\e=0$) the solution for arbitrary $m$ has the form
\bear\label{4.4} %\nonumber
&&\phi= \pm \sr{2m/\k} \ln t +\phi_0,\\
\label{4.4a}
&& V(t)=\k^{-1}(\L+(m+3m^2)/t^2).
\ear
Eliminating $t$ we find an exponential dependence $V$ on $\phi$
\beq\label{4.5}
V(\phi)=\k^{-1}\{\L+(m+3m^2) e^{-\sr{2\k m^{-1}}\{\phi-\phi_0\}}\},
\eeq
what is usually the definition of the power law inflation \cite{olive90}.

In the case of open and closed Universe $(\e \ne 0)$ it is possible to
find an explicit dependence $V$ on $\phi$ just for some values of $m$.
For example, if $m=1$ (in this case the formulas
(\ref{4.2}) and (\ref{4.3}) do not work)
\beq\label{4.6}
V(\phi)=\L/\k +\k^{-1} \exp \{- {2(\phi -\phi_0) \over \pm \sr{2/\k}
\sr{1+\ve K_0^{-2}}}\}.
\eeq
\subsection{Exponential inflation}

The case when the scale factor $K(t)$ of the Universe
grows up very fast by the exponential type law have been analysed in
\cite{barrow94,chezhu96iv}. Let us mention about the differences in
Barrow's and `fine turning of the potential' approaches.
To find new exact solutions, in Barrow's method \cite{barrow94},
one needs to take the scalar field as the function of time $\phi =\phi (t)$
and then
to determine the evolution of the expansion scale factor $K(t)$ and the
potential $V(\phi (t))$ from it. This approach was applied in works
\cite{barpar95},\cite{parbar95} as well.
In `fine turning of the potential' method
 \cite{chezhu95},\cite{chezhu96iv}
the exact solutions have been obtained
by taking, first, the scalar factor as the function of time $K=K(t)$
and then determining the evolution of the potential $V=V(t)$ and the
evolution of a scalar field $\phi = \phi (t)$, the dependence between
$V$ and $\phi$ being, in general, parametric one.
Thus we can combine both of the methods, if we will put the metric
obtained by Barrow's way as the seed solution in `fine turning of the
potential'. As an example of a synthesis of both methods let us consider
one of the exact solutions obtained in \cite{barrow94}.

We can choose the scalar factor of the Universe in the form of the first
class of exact solutions \cite{barrow94}:
\beq\label{4.7} %{3.8}
    K(t)=K_0\cdot{\sinh^{A^2/2}\{2\lambda{t}\}}
\eeq
where $\l$--constant.

This scalar factor (\ref{4.7}) have been obtained for the spatially
flat FRW Universe in \cite{barrow94}. For this solution, using
(\ref{3.6})-(\ref{3.7}), one can find
\bear\nonumber
&&V(t)=\frac{2\e}{K_0^2}\left[\sinh 2\l t\right]^{-A^2}+~~~~~~~~~~~~~~~~~~~~~~~~~~\\
\label{4.8}
&&+\lbrace \L +\l^2A^2 \left[ (3A^2-2)\coth^22\l t +2\right]\rbrace ,\\
\nonumber
&&\phi (t)=\\
\label{4.9}
&&\int\sqrt{2A^2\l^2-\frac{\e}{K_0^2}\left[\sinh (2\l t) \right]^{2-A^2}}
\frac{\pm\sqrt{\frac{2}{\kappa}}dt}{\sinh (2\l t)}.
\ear
It is easy to see from (\ref{4.8}) that $2\l^2A^2 $ plays role an
effective (positive) cosmological constant and when $t \to \infty , V(t)$
will approach to $ V(t)|_{\e = 0} $ for open $(\e=-1)$ or closed $(\e=1)$
universes. The scalar field can be calculated exactly when, for example,
$|A| =\sr{2}$.
\bear\nonumber
&&\phi (t)=\pm\sqrt{\frac{2}{\kappa}}\sr{\frac{A^2}{2}-\frac{\e}{4\l^2K_0^2}}
\ln \tanh (\l t)+ \\
\label{4.10}
&&+\phi_0 = \tilde A \ln \tanh (\l t)+\phi_0.
\ear
The dependence $V=V(\phi)$ can be presented as direct one for open universe
\beq\label{4.11}
V(\phi)=\L +2\l^2 +\cosh^2(\frac{\phi}{\tilde A})\left[ 2\l^2-\frac{2}{K_0^2}
\right].
\eeq
\section{Conformally-flat solutions for nonlinear scalar fields}
Let us consider the case
of the conformally-flat metric
\beq\label{5.1}
dS^2=A\{-(dx^1)^2-(dx^2)^2-(dx^3)^2 +(dx^4)^2\}
\eeq
as an application of the method above. In (\ref{5.1}) $ A=A(x^3,x^4)$.
The analogy of the formulas (\ref{3.6}) and (\ref{3.7}) can be presented
in the form
\bear\label{5.2}
V(\phi (x^3,x^4))=\k^{-1}\frac{A_{33}-A_{44}}{2A^2},\\
\label{5.3}
\phi_3^2 +\phi_4^2 =\k^{-1}\lbrace{ -\frac{A_{33}+A_{44}}{A}
+\frac{3(A_3^2+A_4^2)}{2A^2}\rbrace}.
\ear
In special cases it is possible to reduce solutions
(\ref{5.2}) and (\ref{5.3}) to those ones obtained earlier.
The construction of new solutions is also possible.
\subsection{Cosmological and static solutions}
%Let us consider few examples.
Few examples will show the possibilities of the method.
\begin{itemize}
\item
Let $A=A_0e^{H_0\eta},~\eta=x^4.$ The solution is
\bear\label{5.4}
 &&V(\phi (\eta))=-\frac{H_0^2}{\k A_0}e^{\mp\sr{2\k}\phi},\\
\label{5.5}
 &&\phi=\pm\frac{H_0}{\sr{2 \k}} \eta.
\ear
\item
Let $A=A_0e^{H_0z},~z=x^3.$ Then the solution is
\bear\label{5.6}
 &&V(\phi (z))=\frac{H_0^2}{\k A_0}e^{\mp\sr{2\k}\phi},\\
\label{5.7}
 &&\phi=\pm\frac{H_0}{\sr{2 \k}} z.
\ear

Both cases (\ref{5.4}) and (\ref{5.6}) correspond to Liouville-type
nonlinearity for the $V(\phi)$. These solutions have been obtained in the
work \cite{ivanov80} by suggesting the Liouville form of the potential
 $V(\phi)$.
It is clear that the cosmological solution (\ref{5.4})-(\ref{5.5})
is the same as (\ref{4.6}) for $m=1$, but calculated in the conformal time $\eta $.
\item
Let $A=e^\b,$ where $\b=\half c_1\eta^2+c_2\eta+c_3 $. Then, integrating
(\ref{5.2})-(\ref{5.3}), we can find
\bear\label{5.8}
V(\phi (\eta))=-\k^{-1}e^{-\b}\{c_1+(c_1\eta +c_2)^2\},\\
\nonumber
\phi (\eta)=\pm\frac{\sr{2}}{\sr{\k} c_1}\left( \half \tilde\eta\sr{\tilde\eta^2-c_1}-\right.\\
\label{5.9}
\left. -\half c_1\ln\vert\tilde\eta+\sr{\tilde\eta^2-c_1}\vert \right),\\
\label{5.10}
\tilde\eta=\frac{c_1\eta+c_2}{\sr{2}}.
\ear
\item
Let $A=e^\b,$ where $\b=\half c_1z^2+c_2z+c_3 $. Because of the symmetry of
equations (\ref{5.2})-(\ref{5.3}) by respect of $x^3$ and $x^4$ (up to sign)
one can find the static analogy for the cosmological solutions
(\ref{5.8})-(\ref{5.10}) by virtue of the substitution:
$\eta \to z,~V(\phi(\eta)) \to - V(\phi(z)) $.
\item
Let $A=e^\b,$ where $\b=\frac{a_1}{12}\eta^4 +a_3 .$
Then the potential can be presented in the form
\beq\label{5.11}
V(\phi (\eta))=-\k^{-1}e^{-\b}\{a_1\eta^2+\frac{a_1^2}{9}\eta^6\}.
\eeq
The expression for the scalar field reads
\bear\label{5.11-1}
\nonumber
&&\phi (\eta)=\pm\frac{3}{2a_1\sr{2\k}}\left\{ \half \tilde\eta\sr{\tilde\eta^2-c_1}-\right. \\
&&\left. -\half c_1\ln\vert\tilde\eta+\sr{\tilde\eta^2-c_1}\vert \right\},\\
&&\tilde\eta=\frac{a_1\eta^2}{3}
\ear
%can be determined from
%(\ref{5.9}) by virtue of the substitution:
%$\tilde\eta \to \frac{a_1\eta^2}{3\sr{2}},~c_1 \to a_1.$
\item
Let $A=e^\b,$ where $\b=\frac{a_1}{12}z^4 +a_3 .$
The static analogy can be obtained from the case above by substituting:
$\eta \to z,~V(\phi(\eta)) \to -V(\phi(z)).$

It should be noted here that the list of analytical solutions can be extended
if one can solve (\ref{5.3}) for given gravitational field $A(x^3,x^4)$.
\end{itemize}
\subsection{Solitary wave solutions}
Let $A=A(\th),~\phi=\phi(\th), $ where $~\th=z-u_0\eta $.
That is we are looking for solutions of a solitary wave type.
Equations (\ref{5.2})-(\ref{5.3}) are reduced to
\bear\label{5.12}
V(\phi(\th))=\k^{-1}(1-u_0^2)\frac{A_{\th \th}}{2A^2}\\
\label{5.13}
\phi_\th^2=\k^{-1} \lbrace{ -\frac{A_{\th \th}}{A}
+\frac{3A_\th^2}{2A^2}\rbrace}.
\ear
Using the analogy of equations (\ref{5.12})-(\ref{5.13}) to those ones for
cosmological and static solutions above we can conclude that all solutions
presented in this section can be obtained by virtue of the substitutions:\\
$\eta \to \th ,V(\phi(\eta)) \to \pm (1-u_0^2)^{-1}V(\phi(\th)).$
It is interesting to note that, when $u_0^2=1 $, e.g. $u_0^2$ equal to the
speed of light, the potential disappears.

\subsection{The massive scalar field}
The massive scalar field can be considered as the simplest chaotic
inflationary model \cite{linde90}. The analysis of the chaotic inflation
scenario has been based on the asymptotic solution, when
$ m \ll 1,~\phi \gg 1 $. To find the exact solutions in the framework of
general relativity with the massive scalar field is the long standing
problem. To understand the reason of this fact we can insert the potential
\beq\label{5.14}
V(\phi)=\half m^2 \phi^2
\eeq
in the general formulas (\ref{5.2})-(\ref{5.3}). After simple manipulations
with (\ref{5.2})-(\ref{5.3}), using (\ref{5.14}), it is possible to find
the test equation for exact solutions of self-gravitating massive scalar
field in the conformally-flat spaces (\ref{5.1}) in the form:
\beq\label{5.16}
\frac{1}{2m^2}\lbrace \lbrack\frac{A_{44}}{2A^2}\rbrack_4\rbrace^2+
\frac{A_{44}}{2A^2}\lbrace\frac{3}{2}\frac{A_4^2}{A^2} -\frac{A_{44}}{A}
\rbrace = 0.
\eeq

Unfortunately the solution of the form $A=A_0\eta^{-2}$ leads
to $\phi=const$ and can be identified with solutions for
gravitational vacuum with $V(\phi)=const =\L$.
%Thus it is proved the following \\
%{\bf Proposition:} Any solution of test equation (\ref{5.16}) by respect
%of $A(x^4)$ solves the problem of exact solutions for
%massive scalar field in standard inflationary model.
Any other solution of test equation (\ref{5.16}) by respect
of $A(x^4)$ solves the problem of exact solutions for
massive scalar field in standard inflationary model.

The complications of the third order non-linear differential equation
(\ref{5.16}) are the reason of impossibility to find the exact solutions
in standard inflationary model.
Nevertheless it is possible to calculate by virtue of numerical study
the divergence
an asymptotic and numerical solutions from the exact one.

\section{Conclusions}
This review article are devoted to better understanding the gravitational
field of the
very early Universe. We have presented three most important sources of the
gravitation: self-interacting scalar field, chiral field and gauge field.
The correspondence between all of them are considered.
In this contribution more attention have been payed to nonlinear scalar field
source of gravity. The progress in finding the exact solutions in inflationary
universe is reviewed.
The basic idea of `fine turning of the potential'
method is discussed and computational background is presented in
details. A set
of new exact solutions for standard inflationary model and conformally-flat
space-times are obtained. Special attention payed to relations between
`fine turning of the potential' and Barrow's \cite{barrow94} approaches.
As the example of a synthesis of both methods new exact solution is obtained.

The analogous presentation of chiral and gauge fields will appear in future
publications.

\section*{Acknowledgements}

The author is grateful to J.D.Barrow and D.S.Salopek
for sending preprints on relevant topic.

This work was partly supported by the CCPP `Cosmion' in the project on
CosmoParticle Physics.

\small{

}    %  end `small'

\begin{thebibliography}{99}
%1
\bibitem{guth81}
A.H. Guth, Phys. Rev. D 23 (1981) 347.
% 2
\bibitem{linde90}
A.Linde, Particle Physics and Inflationary Cosmology (Gordon and Breach,
New York, 1990).
%  3
\bibitem{olive90}
K.A. Olive, Phys. Rep. 190 (1990) 307.
%   4
\bibitem{linde82}
A.D.Linde,
%{\it A new inflationary universe scenario: a possible solution
%of the horizon, flatness, homogeneity, isotropy and primordial
%monopole problems}
Phys.Lett. B 108, No.6 (1982) 389.%-393.
%    5
\bibitem{albste82}
A.Albrecht, P.J. Steinhardt,
%{\it Cosmology for Grand Unified Theories with Radiatively
%Induced Symmetry Breaking}
Phys.Rev.Lett. 48, No.17 (1982) 1220.%-1223.
% 6
\bibitem{ch95gc}
S.V.Chervon,
%{\it Chiral nonlinear sigma models and cosmological inflation}
 Gravitation \& Cosmology. Vol.1, No.2 (1995) 91.
% 7
\bibitem{brandenberger89}
Brandenberger R.H. Inflationary Universe Models and Cosmic strings.
 In: Physics of the Early Universe (eds. J.A.Peacock, A.F.Heavens,
A.T.Davies), Proc.of the 36th Scottish Universities Summer School in Physics,
1989, Publ. by the Scottish Universities Summer School in Physics, p.281.
%
\bibitem{svv96iv}
S.V.Chervon, V.K. Shchigolev, V.M.Zhuravlev,
Izv.Vyssh.Ucheb.Zaved. Fiz.,N 2 (1996) 41.%-49.
%
\bibitem{ivanov80}
G.G.Ivanov, Izv.vuzov.Fiz., N 12 (1980) 18.
%
%
\bibitem{bbl86}
J.D.Barrow, A.Burd, D.Lancaster, Class.Quantum Gravity  3
(1986) 551.
%
\bibitem{barsai93}
J.D.Barrow, P.Saich, Class.Quantum Grav.  10 (1993) 279.
%
\bibitem{muslimov90}
A.G.Muslimov, Class.Quantum Grav. 7 (1990) 231.
%
%\bibitem{bos92}
%I.L.Buchbinder, S.D.Odintsov, I.L.Shapiro,
%Effective action in quantum gravity. (IOP Publishing Ltd., 1992).
%
%
\bibitem{ellmad91}
G.Ellis, M.Madsen, Class.Quantum Grav. 8 (1991) 667.
%
\bibitem{barrow94}
J.D.Barrow, Phys.Rev. D49 (1994), 3055.
%
\bibitem{barpar95}
J.D.Barrow, P.Parsons, Phys.Rev. D52 (1995), 5576.
%
\bibitem{parbar95}
P.Parsons, J.D.Barrow, Phys.Rev. D51 (1995), 6757.
%
\bibitem{chezhu95}
S.V.Chervon, V.M.Zhuravlev,
%{\it Exact solutions and fine turning of the potential
%in the cosmological inflationary models}
Foundations of gravitation and cosmology.
Abstracts of the reports at the International School-Seminar,
Odessa, September 4-10, 1995, p.67.
%
\bibitem{chezhu96iv}
S.V.Chervon, V.M.Zhuravlev,
%Exact solutions in cosmological inflationary models,
Izv. Vyssh. Ucheb. Zaved. Fiz.,N 8 (1996) 81.
%
\bibitem{salbon90}
D.S.Salopek, J.R.Bond, Phys.Rev. D42 (1990) 3936.
%
\bibitem{salopek91}
D.S.Salopek, Phys. Rev. D43 (1991) 3214.
%
\bibitem{szc96}
V.K.Shchigolev, V.M.Zhuravlev, S.V.Chervon,
%{\it A new class of inhomogeneous cosmological models with
%Yang-Mills fields}
Pis'ma Zh.Eksp.Theor.Fiz. v.64, No. 2 (1996) 65.
%
\bibitem{ch96el}
S.V.Chervon 
{\it Nonlinear fields in theory of gravitation and cosmology.}
Middle-Volga Scientific Center, Ulyanovsk State University, Ulyanovsk, 
1997.-191p.(in Russian)

{\it Non-linear scalar field coupled to gravity:
new exact solutions for inflation scenarios}
(submitted for publications)
%
\bibitem{ch95iv}
S.V.Chervon
Izv.Vyssh.Ucheb.Zaved. Fiz., No.5 (1995) 114.
%
%
\bibitem {shavil84}
Q.Shafi, A.Vilenkin,
%{\it Inflation with SU(5)}
Phys.Rev.Lett. 52, No.8 (1984) 691.%-694.
% 14
\bibitem{colwei73}
S.Coleman, E.Weinberg, Phys.Rev. D 7 (1973) 1888.
%
\bibitem{star85}
A.A.Starobinsky, Pis'ma Zh.Eksp.Teor.Fiz.  42, N 3 (1985) 124.
%
\bibitem{tvvsj87}        %14
 M.S. Turner, J.V. Villumsen, N. Vittorio, J. Silk  and R. Juszkiewich,
Astroph. J.  323 (1987) 423.
%
\bibitem{sakkhl93}
A.S.Sakharov, M.Yu.Khlopov, Yadernaya Fizika {\bf 56} (1993) 220
%
\end{thebibliography}
\end{document}